\documentclass{mem}
\usepackage{natbib}\usepackage{txfonts}\usepackage{balance}
\usepackage{graphicx}
\usepackage{hyperref}
\idline{99}{99}
\begin{document}
\def\teff{$T\rm_{eff }$}
\def\kms{$\mathrm {km s}^{-1}$}

\title{
Magnetic Fields from Filaments to Cores
}

   \subtitle{}

\author{
P. M.\, Koch\inst{1}, Y.-W.\, Tang\inst{1},
N.L.\, Chapman\inst{2},
A.\, Duarte-Cabral\inst{3},
P.T.P.\, Ho\inst{1,4},
G.\, Novak\inst{2},
N.\, Peretto\inst{3},
Y.-N.\, Su\inst{1},
S.\, Takakuwa\inst{5,1},
\and H.-W.\, Yen\inst{6}
          }

\institute{
Academia Sinica Institute of Astronomy and Astrophysics (ASIAA), Taipei, Taiwan;
\email{pmkoch@asiaa.sinica.edu.tw}
\and
Center for Interdisciplinary Exploration and Research in Astrophysics (CIERA) and Department of Physics \& Astronomy, Northwestern University, USA 
\and
School of Physics \& Astronomy, Cardiff University, Cardiff, UK
\and
East Asian Observatory (EAO), Hilo, Hawaii, USA
\and
Department of Physics and Astronomy, Graduate School of Science and Engineering, Kagoshima University, Kagoshima, Japan
\and
European Southern Observatory (ESO), Garching, Germany
}

\authorrunning{Koch}

\titlerunning{B-fields from Filaments to Cores}

\abstract{

How important is the magnetic (B-) field when compared to gravity and turbulence
in the star-formation process? Does its importance depend on scale and location?
We summarize submm dust polarization observations towards the large filamentary 
infrared dark cloud G34 and towards a dense core in the high-mass star-forming region W51.
We detect B-field orientations that are either perpendicular or parallel to the G34 
filament axis. These B-field orientations further correlate with local velocity gradients. 
Towards three cores in G34 we find a varying importance between B-field, gravity, and 
turbulence that seems to dictate varying types of fragmentation.  At highest resolution
towards the gravity-dominated collapsing core W51 e2 we resolve new B-field features,
such as converging B-field lines and possibly magnetic channels.


\keywords{ISM: individual objects: (G34, W51) -- ISM: magnetic field -- polarization 
}
}
\maketitle{}

\section{Introduction}

The exact role of the magnetic (B-) field in the star-formation process is highly 
debated. Zeeman observations indicate field strengths that can be comparable to 
or even dominate over gravity, turbulence or centrifugal force. Simulations 
predict an evolving role of the B-field with time. 
Where and when exactly does the B-field play which role? We summarize B-field observations
that cover (1) filamentary scales (several pc) in the infrared dark cloud G34.43$+00.24$ (hereafter G34) 
where we trace the large-scale field morphology with a resolution $\theta\sim10\arcsec$ ($\sim0.18$pc)
and (2) smaller denser cores ($\sim 0.05$pc) in the high-mass star-forming region W51 where we resolve
new B-field structures with $\theta\sim0\farcs26$ ($\sim5$mpc).
We trace the plane-of-sky projected B-field morphology via dust polarization. Dust grains
are expected to be aligned with their smaller axes parallel to B-field lines \citep[e.g.,][]{hoang16}.
Emission at (sub-)millimeter wavelengths is, thus, polarized perpendicular to field lines.

\section{Results}

\subsection{Large Scales -- Filament}

\begin{figure}[t!]
\includegraphics[width=6.75cm]{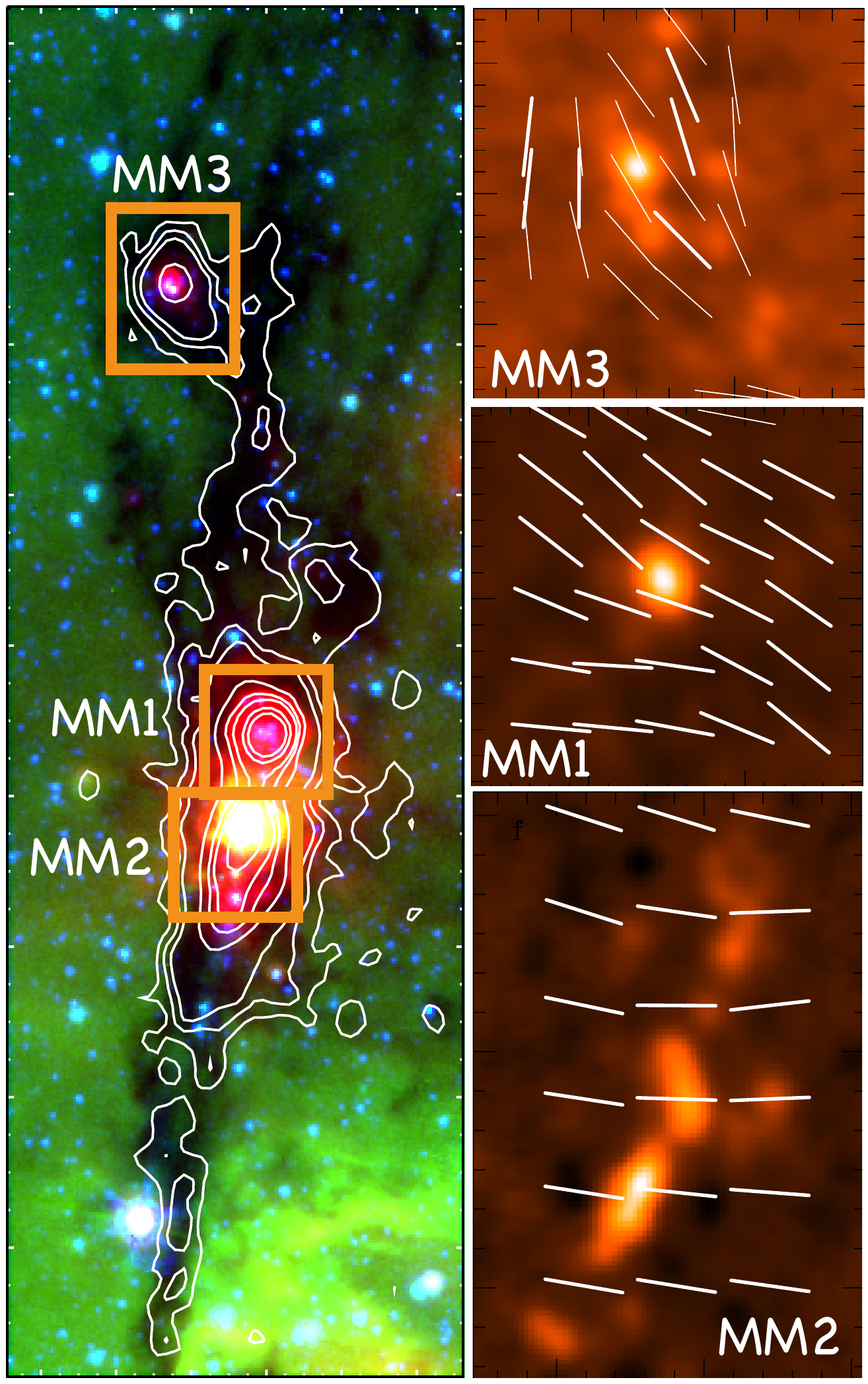}
\caption{\footnotesize
Left panel: G34 three-color composite map from {\it Spitzer} (IRSA archive). Contours are 1mm  
\citep{rathborne06}. B-field segments detected with 
CSO/SHARP at 350$\mu m$ are shown with white segments
in the right panels. Color shows the different fragmentation types, clustered fragmentation in MM3, no 
fragmentation in MM1, aligned fragmentation in MM2, from higher-resolution observations \citep{hull14,zhang14}.
 }
\label{g34}
\end{figure}

The G34 filament extends over about 8pc. 
Its three most massive cores contain several hundred solar masses each within an area of about 0.5pc.
Polarimetric observations with SHARP on the CSO detected the three cores
MM1, MM2, and MM3 with a resolution $\theta\sim10\arcsec$ at a wavelength of 350$\mu m$. 
Polarization was clearly detected along and across the filament \citep{tang17}.
Towards the MM1/MM2 ridge, the B-field is mostly perpendicular to the filament's axis. 
The B-field appears parallel to the filament around MM3 (Fig. \ref{g34}).
Combining N$_2$H$^+$ kinematics (Peretto et al. in preparation) with polarimetric data, we detect 
a very close alignment between local velocity gradients and local B-field orientations across MM1/MM2. This  
is suggestive of the B-field guiding the gas flow.  A detailed analysis of turbulent-to-mean field, B-field 
dispersion and B-field strength, turbulent and B-field pressure, virial parameter and mass-to-flux ratio 
reveals a varying relative importance between B-field (B), gravity (G), and turbulence (Turb) 
in the three cores. We find $\rm G>\rm B>\rm Turb$ in MM1, $\rm B>\rm Turb>\rm G$ in MM2, 
and $\rm G\sim \rm Turb >\rm B$
in MM3 \citep{tang17}. Interestingly, the three cores display also different fragmentation properties on the next smaller
scales (Fig. \ref{g34}). MM1 shows no fragmentation, consistent with our finding that gravity dominates over
both the B-field and turbulence. MM2 displays an aligned fragmentation, likely due to a strong B-field
dictating the locations of the fragments. MM3 reveals a clustered fragmentation resulting from a
significant level of turbulence.
Hence, we propose that the relative importance between B, G, and Turb is responsible
for different fragmentation scenarios. 

\begin{figure*}[h!]
\includegraphics[width=13.5cm]{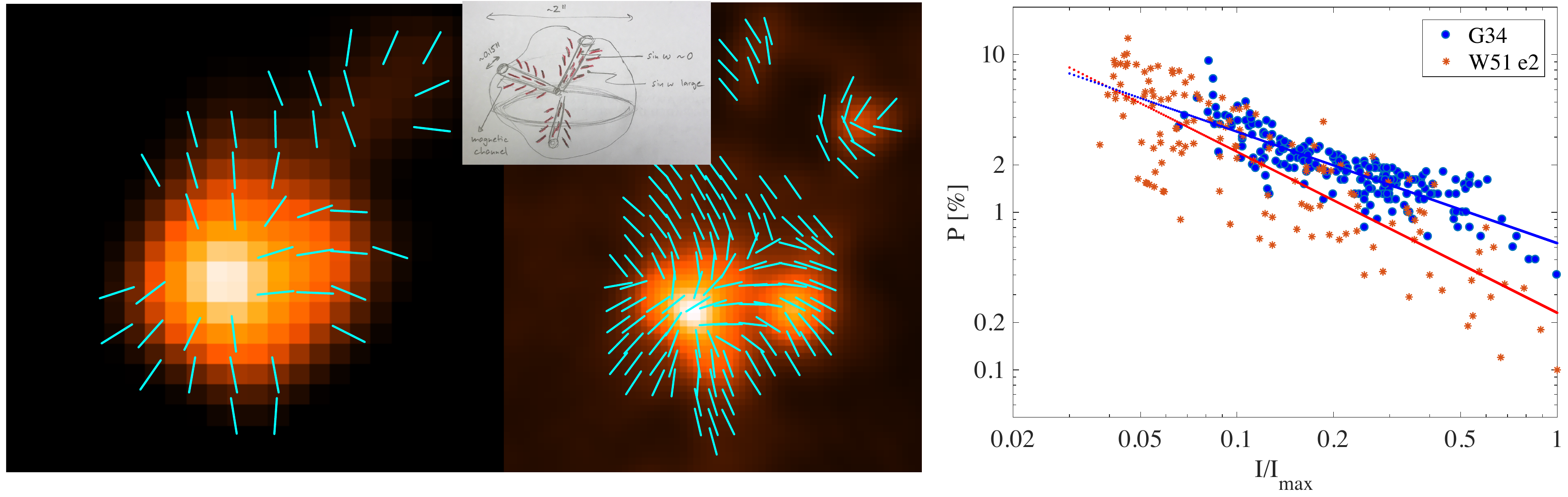}
\caption{\footnotesize
W51 e2 observed with the SMA \citep{tang09} with a resolution $\theta\sim 0\farcs7$ (left panel)
and with ALMA \citep{koch17} with $\theta\sim0\farcs26$ (middle panel).
The color scale is dust continuum. Segments indicate B-field orientations.
The ALMA map reveals new sub-structures such as, e.g., the bow-shock shaped B-field morphology in the 
satellite core in the northwest and the B-field lines symmetrically converging from two sides into a narrow
straight line along a southeast-northwest direction starting from the center.  The inlet is a cartoon of the 
envisioned magnetic channels where material can collapse ($\sin\omega\sim0$) while
outside of the channels the B-field provides maximal resistance against gravity ($\sin\omega$ large). 
Right panel: different slopes, $-0.7$ for G34 and $-1$ for W51 e2, for the intensity-vs-polarization percentage
correlation.
}
\label{w51}
\end{figure*}

\subsection{Small Scales -- Core}

With an improvement in resolution by a factor of about 7 in area over the SMA observations (\citet{tang09};
$\theta\sim0\farcs7$ at 345 GHz), the first ALMA polarization observations towards W51 (\citet{koch17}; $\theta\sim0\farcs26$
at 230 GHz) reveal several new features.  ALMA's sensitivity enables
a complete polarization coverage, clearly detecting and resolving B-field structures in polarization
holes in earlier (less sensitive) SMA data (Fig. \ref{w51}). 
The bow-shock shaped B-field
structure in the northwestern satellite core suggests that this core is falling towards the central more massive core.
Around a southeast-northwest axis starting from the central peak, the field lines appear to symmetrical
converge towards a narrow straight line. This striking symmetry (also observed in several regions in W51 e8 and 
North \citep{koch17} has led to the speculation of B-field convergence zones that result in magnetic channelling (Fig. \ref{w51}).
To further quantitatively assess the role of the B-field on small scales in (collapsing) cores we propose
a new diagnostic: the $\sin\omega$ measure. 
By projecting the direction of the local field tension force onto the direction of gravity,
the effectiveness of the B-field to oppose gravity can be assessed.  The angle $\omega$ between these
two directions is measurable from polarization observations. 
$\sin\omega$ in the range between 0 and 1
measures the fraction of the field tension force that can work against gravity \citep{koch17}.
If the two directions are 
aligned, i.e., $\sin\omega\sim 0$, the B-field cannot slow down gravity. The B-field can maximally oppose gravity when 
the directions are largely misaligned ($\sin\omega\sim1$; Fig. \ref{w51}, inlet).
The possible existence of magnetic channels together with the $\sin\omega$ measure has an interesting application.
One $0\farcs15$ magnetic channel comprises
0.4\% of the entire mass in a $2\arcsec$-diameter sphere.  A network of 10 channels can then reduce the star-formation rate
to 4\%, assuming that the mass inside the channels is converted into stars while the material 
outside is held back by the B-field.

\section{Conclusions}

Combining polarimetric and kinematics data we quantify B-field, gravity, and turbulence towards three
cores in the G34 filament. We conclude that the interplay between these forces varies. Hence, the 
field can play a role from dominant to negligible. The varying relative importance between these forces
seems to be responsible for how fragmentation proceeds. In denser cores in W51 we possibly 
start to resolve B-field convergence zones that lead to narrow magnetic channels where only a small amount 
of gas can collapse. Different slopes in the intensity-vs-polarization percentage correlation, 
$-0.7$ for G34 and $-1$ for W51 e2 (Fig. \ref{w51}, right panel),
hint different dust grain properties.

\bibliographystyle{aa}

\end{document}